\documentstyle[epsf,twoside,a4wide,12pt]{article}
\newcommand{\sF}{S_{F}}
\newcommand{\stF}{S_{F}^T}

\newcommand\D{\displaystyle}

\newcommand\be{\begin{equation}}
\newcommand\ee{\end{equation}}
\newcommand\bano{\begin{eqnarray*}}
\newcommand\eano{\end{eqnarray*}}
\newcommand\ba{\begin{eqnarray}}
\newcommand\ea{\end{eqnarray}}

\newcommand\la{\langle}
\newcommand\r{\rangle}

\input epsf

\begin{document}

\title{\vskip -2cm\hfill
\\
\vskip 2cm
Application of the Heat-Kernel Method to\\ the Constituent Quark Model\\
 at\\ Finite Temperature
\thanks{Supported by the GSI (Darmstadt)}}
\author{Bernd-Jochen Schaefer
\thanks{e-mail address: schaefer@hybrid.tphys.uni-heidelberg.de}
\\
and\\
Hans-J\"urgen Pirner\thanks{e-mail address:
pir@dxnhd1.mpi-hd.mpg.de} 
\\\\
Institut f\"ur Theoretische Physik\\Universit\"at Heidelberg\\
}

%
\vspace{1.0cm}

\maketitle

\begin{abstract}
\setlength{\baselineskip}{18pt}
\noindent
Abstract: The Heat Kernel Method is applied to the
constituent quark model. We calculate  the effect of 
thermal quark fluctuations on the meson action and the 
resulting quark condensate and $\pi \pi$-scattering
amplitude at finite temperature. The quarks produce a
chiral phase transition only by their effect on the
mesonic coupling constants. The 
s-wave isospin zero $\pi \pi$-scattering amplitude
diverges near the phase transition showing the 
necessity for a more sophisticated treatment of
meson fluctuations.
\end{abstract} 
\newpage \parskip12pt


\newpage
\section{Introduction}
The chiral constituent quark model is a mixture of the
potential model  and the
$\sigma$-model which contains spontaneous chiral
symmetry breaking in a dynamical way.
The quarks get their masses via the interaction with the
scalar field, which develops a vacuum expectation value.
In a previous paper \cite{pirn} the thermodynamics of this model
has been investigated, in particular the
restoration of  chiral symmetry with temperature.
It is important to understand the possible evolution of this model
in QCD in order to handle it correctly at finite temperatures.
For high resolution, i.e. short distance processes fundamental QCD
with quarks and gluons  is the most efficient
theory. A typical scale
associated with perturbative QCD is $\Lambda \geq 1-1.5$ GeV.
RHIC and LHC physics for particles with $p_{\perp} \geq \Lambda$
will be dominated by such processes. In nucleus nucleus collisions, however,
scattering at smaller momentum scale will be non negligible. Since the
strength of the QCD interactions increases with decreasing momentum transfer,
characteristic $q \bar q$ bound states will form and influence the dynamics
at larger distances. A typical resolution where these processes start
to become important
is $\Lambda_{\chi SB}=0.5$ GeV \cite{dirk}.
Below this scale the vacuum changes and acquires a quark condensate and/or
meson condensate. The spontaneous 
breaking of chiral symmetry will persist down to $\Lambda \approx
\Lambda_{QCD}$. Around this scale the confining gluon 
configurations make themselves
felt via confining forces.
In the interval $[\Lambda_{QCD} , \Lambda_{\chi SB}]$ the dynamics is
governed by constituent quarks interacting via pions and sigma mesons.
One should keep in mind that the effective couplings of these mesonic
degrees of freedom to themselves and to the quarks are induced by the
underlying physics of the quarks and gluons.
When one immerses the quarks and gluons into a heat bath, it is to
be expected that the induced couplings will be temperature dependent.
In the following paper we calculate the
temperature dependence of the couplings in the linear sigma model.
We apply the Heat Kernel technique to integrate out
the fluctuations of the fermion variables.

We use the constituent quark model
as an effective theory in the region between $\Lambda_{QCD} 
\approx 200$ MeV and
$\Lambda_{\chi SB} \approx 0.5$ GeV with 
constituent quarks and confining gluons, which are
neglected in this paper \cite{mano}.
If the  integration over quark and meson fields is 
done in one step, then one obtains a first order
phase transition contrary to our expectations for the sigma model and QCD
for zero quark masses.
This calculation therefore presents an intermediate step to illustrate
the technique of the heat kernel which will be finally applied in the
context of
the renormalization
group equations to get a more accurate description of the second order
phase transition. In the course of developing this technique we solve the
problem of quark condensate in the constituent quark model. The natural
order parameter in the linear sigma model is the vacuum expectation
value of the scalar field. We show how we can relate the vacuum expectation
value of the scalar field to the quark antiquark condensate.
In addition we calculate $\pi \pi$-scattering as a function of temperature.
This calculation shows the slightly different physics concept underlying the
constituent quark model and the Nambu--Jona-Lasinio (NJL) model in a physical
observable. The $\pi \pi$-scattering amplitude can be tested via
transport properties. We also show that no new renormalization
constants or cut offs are necessary in the constituent model contrary to
the NJL model, where the meson loops generate additional singularities
which have to be regulated by new phenomenological cut off parameters.
The heat kernel technique allows to restrict the effect of quarks 
of a fixed mass to the bosonic vertices. That means we propose to ignore
the free fermion loop contribution of unconfined quarks in the free energy and also 
to freeze the quark mass in order not to allow artificially long distance
meson meson interactions, which would be cut off by the confining 
forces between quarks and antiquarks below the critical temperature. 
In the constituent quark model it is assumed that the meson degrees of freedom
persist beyond the chiral phase transition. Since the gluonic forces are
especially strong in the $\pi$ and $\sigma$ channels  these mesons may not
dissolve but remain active degrees of freedom  beyond the chiral transition.

The outline of the paper is as follows: In the next section 2
we derive 
the heat kernel method at zero temperature and apply it 
to the constituent quark model at finite temperature
in section 3. We calculate the free energy, which is the necessary
thermodynamical tool for the analysis of a phase transition 
and study the chiral
transition. In section 4 we evaluate the quark condensate
and the $\pi \pi$-scattering length as functions of temperature explicitly.
At the end we conclude and compare our results to other calculations.

\section{The Heat Kernel Method for $T=0$}
\label{AIII}
We consider the chiral constituent model with quarks, $\sigma$ and
$\pi$ mesons.
At zero temperature in Euclidean space the partition function
or the generating
functional without external sources $\Delta$ is given by
\ba\label{zz1}
Z[\Delta=0] & = & \int {\cal D}q {\cal D}\bar{q} {\cal D}\sigma  {\cal D}\vec{\pi}
 \exp\{-\int d^4x \left( {\cal L}_F + {\cal L}_B \right) \}
\ea
with a fermionic ${\cal L}_F$ and a bosonic ${\cal L}_B$ part.
\ba\label{fermil}
{\cal L}_F & = & \bar{q}(x) \left( \gamma_E \partial_E +
g\left( \sigma + i\vec{\tau}\vec{\pi}\gamma_5 \right)\right) q(x)
\ea
\ba\label{bosel}
{\cal L}_B & = & \frac{1}{2} \left(
(\partial_\mu \sigma)^2 + (\partial_\mu \vec{\pi})^2 \right) -
\frac{\mu_0^2}{2}(\sigma^2+\vec{\pi}^2)
+\frac{\lambda_0}{4}(\sigma^2+\vec{\pi}^2)^2\ .
\ea
The Yukawa coupling of the constituent quarks to the mesons is
denoted by $g$. 
The parameters of the linear $\sigma$-model
at $T=0$ are chosen in the same way as in ref. \cite{pirn}.
The mass squared  $\mu_0^2$ is positive reflecting
the spontaneously broken ground state.
The minimum of the potential
$U(\sigma)$
lies at  $\langle\sigma_0\rangle=0.093$ GeV.
The light constituent quarks have a mass of
$m=300$ MeV, the $\sigma$-mass $m_{\sigma}=700$MeV,
$\mu^2_0=(0.495$ GeV$)^2,
\lambda=28.33$ and $g=3.23$. These couplings are
effective couplings obtained after the integration over high momentum
quark and gluon degrees of freedom.
Therefore we must not  reintegrate these high momentum modes at 
zero temperature,
since their effect is already included in the couplings. In the
following we demonstrate the heat kernel
technique at zero temperature to set up the 
notation we are going to use at finite temperature.
Also in a later paper we are investigating the renormalization
group flow with the same method \cite{bjhj}.

By means of a formal integration over the quarks 
one obtains a non-local determinant, which
is in general complex
\bano
Z[\Delta=0] & = &  \int {\cal D}\sigma  {\cal D}\vec{\pi}
 \det \left(\gamma_E \partial_E + g\left( \sigma + i\vec{\tau}\vec{\pi}
 \gamma_5 \right)\right) \exp\{-\int d^4x  {\cal L}_B \}
\eano
\bano
\det \left(\gamma_E \partial_E + g\left( \sigma + i\vec{\tau}\vec{\pi}
\gamma_5\right)\right) & = & \int {\cal D}q {\cal D}\bar{q} 
 \exp\{-\int d^4x {\cal L}_F  \}.
\eano
This determinant can be expressed as a logarithm and 
in the following we consider the real part of the effective action
(the modulus of the determinant)
\ba\label{zz3}
\Re e \, \sF & = & -\frac{1}{2} Tr \ln DD^+
\ea
with the elliptic operator
\bano
DD^+ & = & -\partial_E^2 + g^2 MM^{+} + g\gamma_\mu (\partial_\mu 
M^{+})
\eano
and $M(x) =  \sigma (x) + i\vec{\tau}\vec{\pi} (x) \gamma_5$.
The symbol '$Tr$' is defined by a functional trace over an operator $\cal O$
\bano
Tr {\cal O} \equiv \int d^d x\, tr \langle x|{\cal O}|x \rangle
\eano
and $tr$ is reserved for the remaining trace over all inner spaces 
i.e. spin, positive and negative energy solutions
associated with the gamma matrices, isospin and  color \cite{ball}.

Since $DD^+$ is a positive definite operator it is possible to rewrite the 
logarithm as an Schwinger proper time integral. Then we find for the real part
of the effective action in $d=4$ dimensions
\bano
\Re e \, \sF & = & \lim_{\Lambda \to \infty} \frac{1}{2}
\int_{1/ \Lambda^2}^{\infty}  \frac{d\tau}{\tau} \, Tr e^{-\tau D D^+}\\
& = & \lim_{\Lambda \to \infty}  \frac{1}{2} \int_{1/ \Lambda^2}^{\infty}
\frac{d\tau}{\tau}
\int d^4 x H_0 (x,x;\tau)
\eano
where the second line defines the diagonal part of the zero temperature heat kernel
$H_0 (x,x;\tau)$. 
In general the proper time integration diverges for 
vanishing temperature therefore we introduce a proper time regulator $\Lambda$.
The order of magnitude for $\Lambda$ is estimated in ref. 
\cite{eber} from the meson spectrum  
and lies in the range of about 1 GeV. Physically one can interpret $\Lambda$ as a 
scale at which chiral
symmetry breaking starts.

The non-diagonal part of the heat kernel $H_0(x,y;\tau)$  can
further be evaluated by including a plane wave basis \cite{nepo}. We get
\ba\label{heat0}
H_0(x,y;\tau) & = & tr e^{-\tau D D^+_x} \langle x|y\rangle\nonumber\\
& = & tr \int \frac{d^d k}{(2\pi)^d} e^{-\tau D D^+_x} \langle x|k\rangle
\langle k|y\rangle \nonumber\\
& = & tr \int \frac{d^d k}{(2\pi)^d} e^{-iky} e^{-\tau D D^+_x} e^{-ikx}
\ea
which yields the diagonal part ($y \to x$) of the heat kernel
\bano
H_0(x,x;\tau) & = & tr \int \frac{d^d k}{(2\pi)^d}
e^{-\tau \left( k^2 -2ik_\mu \partial_\mu  + D D^+_x \right)} {\bf 1}\ .
\eano
Note, the unit operator on the right is to be 
associated with other possible functions on  which the 
derivatives with respect to $x$ can act. 
The product rule of differentiation can be implemented by the substitution 
$\partial_\mu \rightarrow \partial_\mu + i k_\mu$ in the operator $D D^+_x$. 
After 
a subsequent rescaling of the momenta $k_\mu \rightarrow k_\mu /\sqrt{\tau}$ 
the following expansion in powers of the proper time $\tau$ is found
\ba
\label{x12}
H_0(x,x;\tau) & = & \frac{tr}{(4\pi\tau)^{d/2}} \int \frac{d^d k}{\pi^{d/2}}
e^{- k^2} \sum_{n=0}^\infty \frac{1}{n!}
\left( \sqrt{\tau}2ik_\mu \partial_\mu - \tau D D^+_x \right)^n  {\bf 1}\ .
\ea
The momentum integration can be done using the relation
\ba
\int \frac{d^d k}{ \pi ^{d/2}}
e^{- k^2 } k_{\mu_1}\ k_{\mu_2} \cdots k_{\mu_{2m}} & = &
\frac{1}{2^m} \delta_{\mu_1 \mu_2 \cdots \mu_{2m}} \nonumber\\
& = & \frac{1}{2^m} \sum_{\sigma \in S(2m)}
\delta_{\mu_{\sigma(1)}\mu_{\sigma(2)}}
\delta_{\mu_{\sigma(3)}\mu_{\sigma(4)}}\cdots
\delta_{\mu_{\sigma(2m-1)}\mu_{\sigma(2m)}}\nonumber
\ea
with $\sigma(1) < \sigma(2)\ ,\  \sigma(3) < \sigma(4)\ ,\  \cdots$ and
$\sigma(1) < \sigma(3) < \sigma(5)\  \cdots$\ . All in all there are
$\ (2m-1)!!\ $ permutations. 
For the constituent quark model the corresponding expression for the heat kernel
in $d=4$ dimensions looks like
\bano
H_0(x,x,\tau) & = & \frac{tr}{(4\pi\tau)^2}
\int \frac{d^4 k}{\pi^2} e^{-k^2} e^{-\tau m^2}
e^{-\tau (-\partial^2 -\frac{2 i}{\sqrt{\tau}} k\cdot \partial + \Omega +V
-m^2)} {\bf 1}
\eano 
with $\Omega := g^2 MM^+$ and $V := g\gamma \cdot (\partial M^+ )$.
A mass term $m^2$ in the exponent has been
introduced to guarantee 
the convergence of the proper time integration. It is chosen
equal to the quark constituent mass  in the vacuum.  
Quantum fluctuations of the  theory for short
distances enter in the $\tau \to 0$ region of the $\tau$ 
integration while 
the behaviour of quantum fluctuations at  long distances 
contribute in the limit $\tau \to \infty$. 
Therefore through the introduction of a mass term $m^2$
no infrared divergences for $\tau \to \infty$ emerge. 
The mass term $m^2$ is an IR regulator and
is not uniquely determined. We checked the sensitivity of 
our later results with respect to variations in $m^2$ and found
that in a reasonable interval there is no strong dependence.

The heat kernel method makes the fermion
functional integration doable by expanding the exponential in the proper
time variable up to the second order. 
Performance of the inner traces yields the following result for
the effective action    
\bano
\Re e \, \sF& = & 4 N_c \lim_{\Lambda \to \infty} \int_{1/ \Lambda^2}^{\infty}\frac{d\tau}{\tau}
\frac{1}{(4\pi\tau)^2} e^{-\tau m^2}\nonumber\\
&& \left[ 1-\tau \left\{ g^2(\sigma^2+\vec{\pi}^2)
- m^2 \right\}
+ \frac{\tau^2}{2} \left\{g^4(\sigma^2+\vec{\pi}^2)^2\right.\right.\nonumber\\
&& \left.\left. + g^2((\partial_\mu \sigma)^2
+ (\partial_\mu \vec{\pi})^2)+ m^4 -2m^2 g^2 (\sigma^2+\vec{\pi}^2) +
g^2 \partial^2
(\sigma^2+\vec{\pi}^2)\right\} \right]\ .
\eano
The proper time integration leads to an incomplete gamma functions 
in the fermionic action
\bano
\Re e \, \sF & = & \frac{m^4}{4\pi^2}\sum_{k=0}^\infty 
\frac{\Gamma (k-2,\frac{m^2}
{\Lambda^2} )}{m^{2k}} Tr\, h_k
\eano
where $Tr\, h_k$ symbolizes a short hand notation for 
the integrals resulting from different powers in $\tau$
containing the fields and their derivatives \cite{eber}.
One observes that the proper time expansion 
leads to a gradient expansion,
since the proper time has dimension $mass^{-2}$. The first
three coefficients of this expansion are always divergent 
for $\Lambda \to \infty$, 
therefore a regularization scheme should be applied at this stage.

\section{The Heat Kernel Method at finite Temperature}
\label{AII}
Before we apply the heat kernel method 
to the constituent quark model we want to show how
to extend it to finite temperature.

The main difference in finite temperature field theory compared 
to zero temperature field theory is the change of the 
continuous time integration of the effective action to a finite integration in
Euclidean space from zero to $\beta = 1/T$. We can generalize
the definition of the heat kernel at zero
temperature cf. Eq. (\ref{heat0})  to finite
temperature
\begin{eqnarray}
H_{T} (x,x',\tau) &=& e^{-\tau DD^{+}} \langle x|x'\rangle_\beta \nonumber \\
&=& e^{-\tau DD^{+}} \delta_{\beta}^4 (x-x')
\end{eqnarray}
where $'\beta'$ stands for the effect of finite
temperature in the Fourier representation of $\delta_{\beta}^4 (x-x')$. 
This means 
one has to insert the correct
boundary condition of the fermions (or bosons) into the Dirac delta function.
For the quarks we have to take anti-periodic boundary conditions in Euclidean
time, i.e. the Fourier representation has only half integer Matsubara
frequencies.
With the help of a generalized theta-transformation one can 
find a connection between
the zero temperature 
and the finite temperature heat kernel for fermions \cite{bosc}
\be\label{hkferm}
H_{T} (x,x,\tau) = H_0 (x,x,\tau) \left[ 1+ 2 \sum_{n=1}^\infty (-1)^n
e^{-n^2 \beta^2 / 4\tau}  \right]\ .
\ee 
This result is remarkable, because the finite temperature heat kernel
separates into a sum  of the zero temperature heat kernel and
a temperature dependent piece, which again is a product of the
zero temperature heat kernel times a temperature dependent function.

\begin{figure}[hbt]
\unitlength1cm
\begin{center}
\begin{picture}(15,9)
\put(9.0,7.3){$\mu ' /10$ [MeV]}
\put(3.1,5.3){$\lambda '$}
\put(3,2.2){$10 z$}
\put(5.5,-0.5){Temperature [MeV]}
\epsfbox{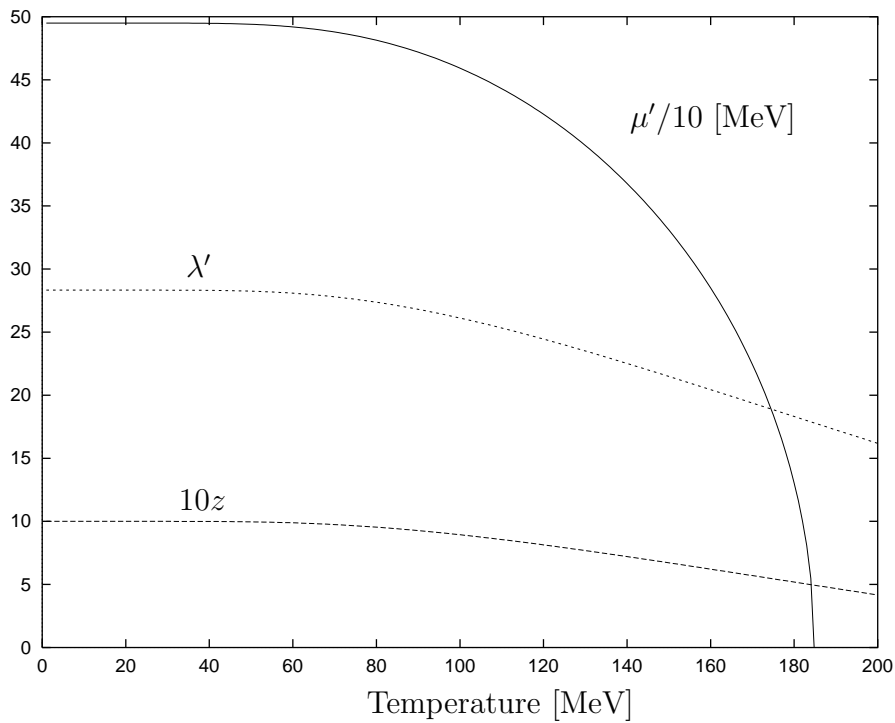}
\end{picture}
\end{center}
\caption{\label{fig1} couplings $\lambda '$, $\mu '$ and $z$
as function of the temperature.}
\end{figure}

Since we are interested in the pure temperature contribution of the
effective action we consider only the difference between the finite temperature
and zero-temperature heat kernel. 
The temperature dependent effective action is 
then given by
\bano
\Re e \, \stF & = & \frac{1}{2} \int d^4 x \int_0^\infty \frac{d\tau}{\tau}
\left\{ H_T (x,x;\tau) - H_0 (x,x;\tau) \right\}\ .
\eano
The vacuum divergences in both kernels cancel each other and no new divergences
are generated at finite temperatures. Therefore we do not need the regulator 
$\Lambda$ anymore and take the proper time integration over all 
positive real $\tau$.   
With the result of Eq. (\ref{hkferm}) we find
\bano
\Re e \, \stF & = & \frac{1}{2} \int d^4x \int_0^\infty \frac{d\tau}{\tau}
2 H_0 (x,x;\tau) \sum_{n=1}^\infty (-1)^n e^{-n^2 \beta^2 / 4\tau}\ .
\eano
In the previous section the expansion of $H_0 (x,x;\tau)$ was shown and by using the 
formula
\be
\int_0^\infty d\tau \tau^{\nu -1} e^{-\frac{\alpha}{\tau} - \gamma \tau}
= 2 \left( \frac{\alpha}{\gamma} \right)^{\nu /2} K_\nu (2 \sqrt{
\alpha \gamma } )
\ee
where $K_\nu (x)$ denotes modified Bessel functions, we finally obtain
\bano
\Re e \, \stF & = & \int d^4 x \left[
\frac{N_c}{2\pi^2} \frac{8m^2
}{\beta^2} F_2 (\beta m) -
\frac{N_c}{2\pi^2} \frac{4 m}{\beta} F_1 (\beta m) \{ g^2 ( \sigma ^2 +\vec{\pi}^2)
-m^2 \} \right.\\
&& + \frac{N_c}{2\pi^2} F_0 (\beta m) \left\{ m^4 + g^2 ((\partial_\mu \sigma )^2
+(\partial_\mu \vec{\pi} )^2) + g^4 ( \sigma ^2 +\vec{\pi}^2)^2\right. \\
&& \left.\left.- 2 m^2 g^2 ( \sigma ^2 +\vec{\pi}^2)
+\frac{1}{3} g^2 \partial^2 ( \sigma ^2 +\vec{\pi}^2) \right\}\right]\ .
\eano   
Here we  define 
\bano
F_i (\beta m) & := & \sum_{n=1}^\infty (-1)^n \frac{K_i (n\beta m)}{n^i}
\qquad \mbox{for}
\ \ i = 0,1,2\ .
\eano
This result is substituted into the partition function Eq. (\ref{zz1}), which 
then becomes only a functional of the mesonic degrees of freedom alone.
\be\label{zmeson}
\D {Z}  = \int D\sigma D\vec{\pi} \exp \left( - \frac{8N_c}{(4\pi)^2} \left\{
\frac{8 m^2}{\beta ^2} F_2 \left(\beta m\right)
+ \frac{4 m^3}{\beta} F_1 \left(\beta m\right) + m^4 F_0
\left(\beta m\right)\right\} \right) \times
\ee
\bano
&\D \exp \left( - {\int d^4 x \left[ \frac{1}{2} z(T) \{
(\partial_\mu \sigma )^2 +(\partial_\mu \vec{\pi} )^2) \}
- \frac{ {\mu '}^2 (T)}{2} ( \sigma ^2 +\vec{\pi}^2 ) +
\frac{\lambda ' (T)}{4} ( \sigma ^2 +\vec{\pi}^2)^2 \right] } \right)&\nonumber
\eano
The fermionic integration has given a prefactor which comes from 
the quark loop at finite temperature. This prefactor is due to the
free motion of quarks, which we think is inadequately described
by the linear sigma model without confinement. Therefore we will
not include it in the  equation of state. We take, however, 
the fermionic effects on the mesonic couplings  seriously.
Due to the freezing of the infrared parameter $m$, the fermion
loop contributions
arise  from short time fluctuations in the thermal system, where
quark confinement is not relevant. 
After sorting one encounters the following temperature dependent 
wave function renormalization constant and 
couplings
up to second order in the
proper time expansion:
\ba\label{Zf}
z (T) & = & 1 + \frac{8 N_c}{(4\pi)^2} 2 g^2 F_0 \left(\beta m\right) \\
\nonumber\\
\label{muf}
{\mu'}^2 (T) & = & \mu^2_0 + 
\frac{8 N_c}{(4\pi)^2} 4 m^2 g^2 F_0 \left(\beta m\right)
+\frac{8 N_c}{(4\pi)^2} \frac{8 m}{\beta} g^2 F_1 \left(\beta m\right) \\
\nonumber\\
\label{lamf}
\lambda ' (T) & = & \lambda_0 + 
\frac{8 N_c}{(4\pi)^2} 4 g^4  F_0 \left(\beta m\right)\ .
\ea
The first terms on the right hand sides in Eqns. (\ref{Zf}-\ref{lamf}) 
are the $T=0$ terms
coming from the bosonic action Eq. (\ref{bosel}).
The finite temperature corrections are generated from the heat kernel.
Since the heat kernel expansion is an one loop integration, 
its result corresponds to the results in perturbation theory
which gives 
temperature dependent modifications of the couplings. The wave function 
renormalization comes from the graph where a quark radiates a meson.
The mass of the mesons is changed by the meson polarization operator in
quark antiquark pairs to order $g^2$. The meson meson scattering 
amplitude gets modifications from the box graph in fourth order $g^4$ of the
coupling.
The numerical values of the couplings are shown in Fig. \ref{fig1} as 
a function of temperature. One sees that $\mu'$ decreases and becomes
zero at $T \approx  185 $ MeV. The coupling $\lambda'$ and the wave
function renormalization $z$ also become smaller. Note the
wave function renormalization $z$ does not vanish at the chiral
transition, which means the pion has not dissolved at this temperature.
The probability to find a free pion at $T\approx 185$ MeV, however, has
decreased to $50 \%$.

We also apply the presented heat kernel method to evaluate the quark
condensate $\la \bar{q} q \r$ at finite temperature in the chiral limit
and to calculate $\pi \pi$-scattering for finite pion masses.
These calculations are possible by adding  an additional source term 
\bano
2 \Delta  \bar{q} q & = & \Delta \sum_{i=1}^2 \bar{q_i} q_i
\eano
to the fermionic Lagrangian Eq.~(\ref{fermil}) and 
then differentiating with respect to $\Delta$.
Thus the parameter $\Delta$ plays the role of a current quark mass
and breaks chiral
symmetry explicitly. We again calculate only the difference 
between the finite temperature heat kernel $H_T$ and
the zero temperature $H_0$. 
Therefore we have to include in 
our bosonic action at zero temperature 
an
explicit symmetry breaking term, arising from the 
integration over the quarks with finite masses at zero temperature
which we do not execute explicitly. 
This term will receive a temperature dependent modification
in our calculation. In addition another term cubic in
meson fields will arise at
finite temperature through the fermion integral.
The symmetry breaking
term in the linear sigma model at $T=0$ is given by ${\cal L}_c = - c \sigma$ 
with a value of $c$ which reproduces the physical pion mass.
\bano
c \sigma & = & - m_\pi^2 f_\pi \sigma .
\eano
Using the 
relation $m_\pi^2 f_\pi^2 = - 2\Delta \la \bar{q}q \r_0$ we get
\bano
c =  \Delta \left( \frac{2\la \bar{q}q \r_0}{f_\pi}\right)  =: 
\Delta \tilde{c}   
\eano
Employing the heat kernel technique we obtain the final effective 
bosonic action at $T \neq 0$ including explicit symmetry breaking
due to finite quark masses $\Delta$.
The meson wave function renormalization constant  has been scaled out.
\bano
S[\sigma , \vec{\pi} ] & = & \int d^4 x\left\{ const + \frac{1}{2}
\left( \left( \partial_\mu \sigma \right)^2 + 
\left( \partial_\mu \vec{\pi} \right)^2 \right)-
\frac{\mu''(T)^2}{2} 
\left( \sigma^2 + \vec{\pi}^2 \right) +\right.\\
&&\left.+ t_3(T) \Delta \sigma  \left( \sigma^2 + \vec{\pi}^2 \right) +
\frac{\lambda''(T)}{4} 
\left( \sigma^2 + \vec{\pi}^2 \right)^2+
t_1(T) \Delta 
\sigma\right\}
\eano
with additional  temperature dependent functions
entering the symmetry breaking terms.
\ba\label{star}
t_1(T) & = & \frac{- 2 (\mu'(T)^2-\mu_0^2)
+ \tilde{c}}{z(T)^{1/2}}\nonumber\\
t_3(T) & = & \frac {4 (\lambda'(T)-\lambda_0)}{z(T)^{3/2}}\nonumber\\
\mu''(T)^2 & = & \frac{\mu'(T)^2}{z(T)}\nonumber\\
\lambda''(T)& = & \frac{\lambda'(T)}{z(T)^2}
\ea
There are two new temperature dependent loop corrections in this formula 
proportional to the external source $\Delta$.
The sigma tadpole term $t_1 \Delta
\sigma$ gets a temperature dependent fermion loop correction.
A temperature dependent
Yukawa vertex between two mesons and the sigma meson is generated due
to the quark loop. The term $t_3 \Delta $ 
adds to the Yukawa vertex produced after
spontaneous symmetry breaking, when $\sigma$ develops a vacuum expectation
value.   
In Table 1 , we give the numerical values of $\mu'',\lambda'',
{t_1 \Delta}/{(f_{\pi} m_{\pi}^2)}$ and $t_3 \Delta$.
Both terms depending on the explicit symmetry breaking $\Delta$ influence the
order of the phase transition. Previously \cite{hild} 
we did not allow such temperature dependent 
terms in calculations of the order of the phase transition. Especially
a third order term would  add to the determinant
term in Flavour-$SU(3)$ from instantons which is nine times larger at $T_c$ 
\cite{hmbj}.

\begin{table}
\begin{center}
\begin{tabular}{|c||c|c|c|c|c|c|}
\hline
$T$ [MeV]        & 0 & 40. & 80. & 120. & 160. & $T_c \approx$ 184.8\\
\hline\hline
$\mu ''$ [MeV]   & 495. & 495. & 493. & 469. & 362. & 0. \\ \hline
$\lambda ''$ [1] & 28.3 & 28.4 & 30.1 & 36.9 & 53.0 & 72.9\\ \hline
$\Delta t_1/ f_\pi m_\pi^2$ [1]     & -1.  & -1.  & -0.97& -0.81& -0.42& 0. \\ \hline
$\Delta t_3$ [MeV]     & 0.   & -0.1 & -7.4 & -38.2& -116.4& -219. \\\hline
\end{tabular}
\end{center}
\caption{\label{tab1} Couplings in the effective Boson action $S[\sigma , \vec{\pi}]$
for different temperatures.}
\end{table}

\section{Quark Condensate and $\pi \pi$-Scattering}

\begin{figure}[hbt]
\unitlength1cm
\begin{center}
\begin{picture}(15,9)
\put(10.1,6.3){mean-field}
\put(11.1,6.1){\vector(-1,-1){0.6}}
\put(5.9,4.7){with bosonic}
\put(8.8,4.5){\vector(1,0){0.7}}
\put(5.9,4.2){fluctuations}
\put(5.5,-0.5){Temperature [MeV]}
\epsfbox{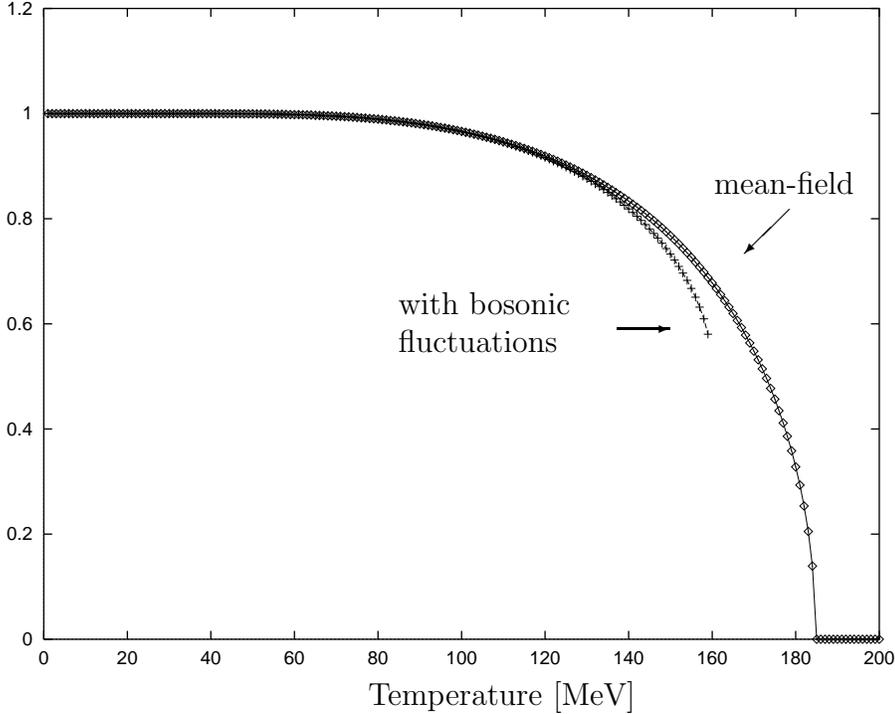}
\end{picture}
\end{center}
\caption{\label{fig2} Normalized condensate $\la\bar qq\r_{MF}$ and $\la\bar qq\r$
with bosonic fluctuations as function of the temperature.}
\end{figure}

The next task is to track the behaviour of the
vacuum expectation value of the $\sigma$ field.  
When we expand the effective action around the spontaneously broken 
ground state
$\sigma_0 = const$, we get 
up to a normalization factor $\cal N$  
the partition function with
the tree contribution to the effective action.
This action contains all the effects of the  temperature fluctuations
of the fermions besides the free energy of free quarks and
antiquarks. These effects suffice to produce a chiral phase transition. 
\bano
S[\sigma_0] & = & (\beta V) (t_1 \Delta \sigma_0 - \frac {\mu''^2}{2}
\sigma_0^2 + t_3 \Delta \sigma_0^3 +  \frac{\lambda''}{4} 
\sigma_0^4)
\eano
In mean field approximation we determine from 
this tree action the quark condensate in the chiral limit as
a function of temperature.
\ba
2\la \bar{q} q \r_{MF} & = & -\frac{1}{\beta V} 
\left.\frac{\partial \ln Z_{MF}(\Delta)}{\partial \Delta}
\right|_{\Delta = 0}= t_1 \sigma_0  + t_3 \sigma_0^3 .
\ea
Minimizing the action with respect to $\sigma$ one finds that
$\sigma_0$ and the resulting quark condensate are smooth
functions of temperature (cf. Fig. \ref{fig2}), 
which vanish at the chiral transition temperature 
where $\mu''^2$ vanishes.
The phase transition is at a higher temperature compared to
the calculation in ref. \cite{pirn}, because the effects 
of the quark fields enter only via the intermediate
coupling of the meson fields. If one would set $m=g \sigma_ 0$
in the prefactor in Eq. (\ref{zmeson}), a variation of this part would give 
a faster decrease of the vacuum expectation value of $\sigma_0$.
But this procedure is questionable since the quarks are not
free below the transition temperature.

In order to estimate the mesonic effects
on the quark condensate we include their thermal fluctuations
in the partition functions.
In the linear sigma model the integration over the bosonic
degrees of freedom is much simpler than in the NJL model \cite{flor}.
The zero temperature bose fluctuations are already
included in the meson coupling constants at zero temperature.
Therefore we use 
a simple one loop approximation for the boson integration.
Including 
mesonic terms 
we obtain 
the following partition function, where the integrals are to be understood at
finite temperature with the mean field masses.
\bano
Z & = & {\cal N} e^{\D -S[\sigma_0] -\frac{1}{2} Tr\ln (-\partial^2 + M^2_\sigma )
-\frac{3}{2} Tr\ln (-\partial^2 + M^2_\pi )}
\eano
\ba\label{masspi}
M^2_\sigma & = & -\mu''^2  + 3\lambda'' \sigma_0^2 
+6 t_3 \Delta \sigma_0\nonumber\\
M^2_\pi & = & -\mu''^2 + \lambda'' \sigma_0^2
+ 2 t_3 \Delta \sigma_0
\ea
The value $\sigma_0$ entering these equations is determined in
mean field from the minimum of the bosonic tree action.
Consequently, $\sigma_0$ depends on the explicit symmetry breaking parameter $\Delta$.
The quark condensate in this approximation contains 
fermionic and mesonic effects. If the couplings $t_1$ and $t_3$ would
not depend on temperature, the quark dynamics would be
invisible in the variation of the quark condensate with temperature.
\ba\label{zweis}
2\la \bar{q} q \r_T = t_1(T) \sigma_0  + t_3(T) \sigma_0^3
+ I_{\sigma} (\sigma_0, T) + I_\pi (\sigma_0, T)
\ea
where the bosonic integrals
are defined as follows:
\bano
I_{\sigma} (\sigma_0, T) & = & -\frac{1}{2} \int \frac{d^3 p}{(2\pi)^3}
\frac{1}{\sqrt{\vec{p}^2 + M_\sigma^2}}
\frac{1}{e^{\beta \sqrt{\vec{p}^2 + M_\sigma^2 }} -1 }
\left.\frac{\partial M_\sigma^2}{\partial \Delta}\right|_{\Delta = 0}\\
& = & - \frac{M_\sigma T}{4\pi^2}
\left.\frac{\partial M_\sigma^2}{\partial \Delta}\right|_{\Delta = 0}  
\sum_{n=1}^\infty \frac{K_1 (n M_\sigma / T)}{n}
\eano
and 
\bano
I_{\pi} (\sigma_0, T) & = & -\frac{3}{2} \int \frac{d^3 p}{(2\pi)^3}
\frac{1}{\sqrt{\vec{p}^2 + M_\pi^2}}
\frac{1}{e^{\beta \sqrt{\vec{p}^2 + M_\pi^2 }} -1 }
\left.\frac{\partial M_\pi^2}{\partial \Delta}\right|_{\Delta = 0}\\
& = & - \frac{T^2}{8}
\left.\frac{\partial M_\pi^2}{\partial \Delta}\right|_{\Delta = 0}\ ,
\eano 
In the last equation we have to take into account
the pion with zero mass.
To specify the result in the chiral limit 
we list the masses and $\sigma_0$ field derivatives
\ba\label{oben}
\left.\frac{\partial M_\pi^2}{\partial \Delta}\right|_{\Delta = 0}
& = & 2 t_3 \sigma_0 + 
( 2 \lambda'' \sigma_0 ) 
\left.\frac{\partial \sigma_0}{\partial \Delta}\right|_{\Delta = 0} \nonumber\\
\left.\frac{\partial M_\sigma^2}{\partial \Delta}\right|_{\Delta = 0}
& = & 3 \left.\frac{\partial M_\pi^2}{\partial \Delta}\right|_{\Delta = 0}
\nonumber\\
\left.\frac{\partial \sigma_0}{\partial \Delta}\right|_{\Delta = 0}
& = &
- \left.\frac{ t_1+3 t_3\sigma_0^2}
{M_\sigma^2}\right|_{\Delta = 0}\ .
\ea


In Fig. 2 we show the resulting quark condensate including the 
bosonic thermal fluctuations. It decreases faster and leads to a first order
transition.
We are aware that the meson one loop approximation is not sufficient 
to generate a second order phase transition for a 
pure mesonic $O(4)$-theory.
In the present paper we are mainly
interested in the effect of quarks on the mesonic action,
in a further paper we are going to improve our calculation by combining 
renormalization group methods with the heat kernel technique \cite{bjhj}.

\begin{figure}[hbt]
\unitlength1cm
\begin{center}
\begin{picture}(15,9)
\put(9.2,6){$T_c$}
\put(9,6.1){\vector(-1,0){0.5}}
\put(5.5,-0.5){Temperature [MeV]}
\epsfbox{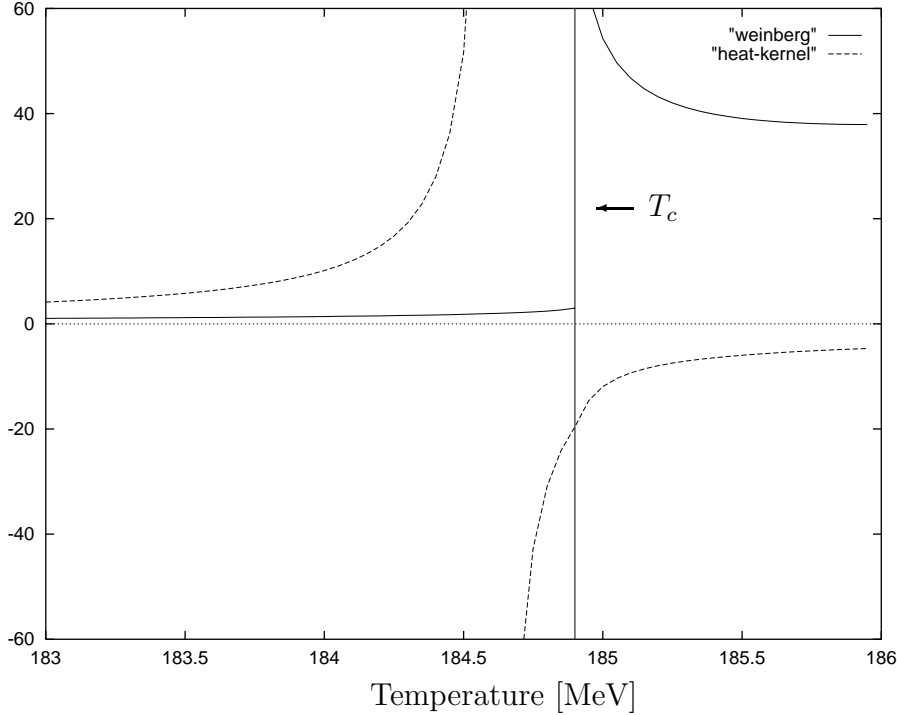}
\end{picture}
\end{center}
\caption{\label{fig3} Scattering lengths $a_{00}$
as functions of the temperature.}
\end{figure}

The $\pi \pi$ s-wave scattering amplitude vanishes in the zero momentum
limit for massless pions. Therefore we have to use the masses
of $\sigma$- and $\pi$-mesons 
in the form with finite symmetry breaking $\Delta$ ( cf. Eq. (\ref{masspi})). 
In addition we need
the Yukawa coupling  $g_{\sigma \pi \pi}$ from the bosonic
action after symmetry breaking:
\be
g_{\sigma \pi \pi}=\lambda'' \sigma_0 + t_3 \Delta
\ee
The s-wave projection of the $\pi \pi$-scattering amplitude in the
isospin zero channel then has the following form
\be
T_{00}(s=4M_{\pi}^2)=\frac{1}{3}\left( (30 \lambda''-4 g_{\sigma \pi \pi}^2
(\frac{9}{M_{\sigma}^2-4M_{\pi}^2} +\frac{6}{M_{\sigma}^2}) \right)
\ee
Inserting the temperature dependent masses Eq.(\ref{masspi}) and coupling constants
Eq.(\ref{star}) we find a divergence of the s-wave isospin zero
scattering length (cf. ref. \cite{huef}) shortly before the chiral phase transition. 
As long as one treats the bosons very naively in mean field or
one loop approximation there comes a point where the $\sigma$
mass becomes degenerate with twice the pion mass, which leads to
an exploding scattering length $a_{00}$
\be
a_{00}=-\frac{ T_{00}}{32 \pi M_{\pi}}
\ee
In Fig. \ref{fig3} we plot the result of our calculation near $T_c$ and compare the 
result with a simple extrapolation of the Weinberg scattering length
\bano
a_{00} & = & - \frac{7 m_\pi}{32 \pi f_\pi^2}
\eano
to finite temperature.

A more careful investigation of pion scattering at finite energies 
shows that the effect of 
the singularity is diminished when one averages over the cm-momentum
of the pions, as the heat bath only contains finite momentum pions.
The total effect is almost completely washed out, if one includes
an average over the unitarized s-wave amplitude, which contains
$\pi \pi$-rescattering effects. It is therefore very important
to include the ring sum or an equivalent damping mechanism to 
control the pion interactions near $T_c$. The scattering length
is a good illustration for the inadequacy of the mean field or
one loop approximation in the presence of Goldstone Bosons.
It remains to be seen how far away from $T_c$ these effects are important
in the order parameter, since in the physical mass case the second
order phase transition is smeared out. 
Also a quantitative comparison of the ring or large $N_f$ approximation
with the renormalization group equations is interesting.
The heat kernel technique allows without any additional assumption
about the form of the infrared regulator to calculate the evolution
equations.
At lower temperatures the presented coupling constants can be included in
the constituent quark model calculation with confinement in 
ref. \cite{pirn} to take into account the effects of thermal quark fluctuations
on the couplings of the sigma model. 
The simplicity of NJL model estimates remains unchallenged by the
linear sigma model. If one wants to tackle higher order effects, however,
our calculation shows that the constituent quark
model is by far superior in its ability to
include meson dynamics. The physical question whether mesonic degrees
of freedom are modeled adequately by this model near $T_c$ has to be
answered by calculating susceptibilities not only in the scalar
and pseudoscalar, but also in the vector channels. These 
results can be compared with numerical lattice calculations but also with 
stunning experimental data from dilepton spectra \cite{dree}.

\section*{Acknowledgments}
We would like to thank J. H\"ufner, D.-U. Jungnickel and
S.P. Klevansky
for useful discussions.

\end{document}